%% file: samplepaper.tex
\documentclass[runningheads]{llncs}
\usepackage[T1]{fontenc}
\usepackage{graphicx}
\usepackage{url}
\usepackage{pifont}
\usepackage{tikz}
\usepackage{comment}
\usepackage{listings}
\usepackage{xcolor}
\usepackage[numbers]{natbib}
\usepackage{xurl}
\usepackage{microtype}
\usepackage{rotating}
\usepackage{marvosym}

\lstdefinelanguage{json}{
    basicstyle=\ttfamily,
    numbers=left,
    numberstyle=\tiny\color{gray},
    stepnumber=1,
    numbersep=5pt,
    showstringspaces=false,
    breaklines=true,
    frame=lines,
    backgroundcolor=\color{white},
    literate=
     *{0}{{{\color{blue}0}}}{1}
      {1}{{{\color{blue}1}}}{1}
      {2}{{{\color{blue}2}}}{1}
      {3}{{{\color{blue}3}}}{1}
      {4}{{{\color{blue}4}}}{1}
      {5}{{{\color{blue}5}}}{1}
      {6}{{{\color{blue}6}}}{1}
      {7}{{{\color{blue}7}}}{1}
      {8}{{{\color{blue}8}}}{1}
      {9}{{{\color{blue}9}}}{1}
      {:}{{{\color{punct}{:}}}}{1}
      {,}{{{\color{punct}{,}}}}{1}
      {\{}{{{\color{delim}{\{}}}}{1}
      {\}}{{{\color{delim}{\}}}}}{1}
      {[}{{{\color{delim}{[}}}}{1}
      {]}{{{\color{delim}{]}}}}{1},
}

\definecolor{delim}{RGB}{20,105,176}
\definecolor{punct}{RGB}{0,0,0}

\def\checkmark{\tikz\fill[scale=0.4](0,.35) -- (.25,0) -- (1,.7) -- (.25,.15) -- cycle;}
\newcommand{\xmark}{\ding{56}}

\newcommand{\redact}[1]{\textcolor{black}{\rule[.5ex]{2em}{0.6pt}}}

\begin{document}
\title{On the Security and Privacy of AI-based Mobile Health Chatbots}

\author{Samuel Wairimu\Letter\inst{1}\orcidID{0000-0003-1750-649X} \and
Leonardo Horn Iwaya\inst{1}\orcidID{0000-0001-9005-0543}}
\authorrunning{Wairimu and Iwaya}
%
\institute{Karlstad University, Universitetsgatan 2, 651 88 Karlstad, Sweden\\
\email{\{samuel.wairimu, leonardo.iwaya\}@kau.se}}

%
%

%

\maketitle              
\begin{abstract}
The rise of Artificial Intelligence (AI) has impacted the development of mobile health (mHealth) apps, most notably with the advent of AI-based chatbots used as ubiquitous ``companions'' for various services, from fitness to mental health assistants. While these mHealth chatbots offer clear benefits, such as personalized health information and predictive diagnoses, they also raise significant concerns regarding security and privacy. This study empirically assesses 16 AI-based mHealth chatbots identified from the Google Play Store. The empirical assessment follows a three-phase approach (manual inspection, static code analysis, and dynamic analysis) to evaluate technical robustness and how design and implementation choices impact end users. Our findings revealed security vulnerabilities (e.g., enabling Remote WebView debugging), privacy issues, and non-compliance with Google Play policies (e.g., failure to provide publicly accessible privacy policies). Based on our findings, we offer several recommendations to enhance the security and privacy of mHealth chatbots. These recommendations focus on improving data handling processes, disclosure, and user security. Therefore, this work also seeks to support mHealth developers and security/privacy engineers in designing more transparent, privacy-friendly, and secure mHealth chatbots.

\keywords{Mobile Health \and mHealth \and Chatbots \and Mobile Apps \and Artificial Intelligence \and Privacy \and Data Protection \and Security}
\end{abstract}
\input{Sections/introduction}

\input{Sections/related_work}

\input{Sections/methodology}

\input{Sections/results}

\input{Sections/discussion}

\input{Sections/limitations}

\input{Sections/conclusion}

\input{Sections/credits}

\bibliographystyle{splncs04}
\bibliography{main}

\end{document}

%% file: Sections/introduction.tex
\section{Introduction}

In recent years, the field of mobile health (mHealth) apps has seen substantial advances with the integration of Artificial Intelligence (AI), particularly generative AI and large language models (LLMs). Several companies are developing mHealth chatbots that leverage LLMs, enabling users to receive preliminary diagnostics and personalized healthcare information \cite{surani2022understanding}. Chatbots, which have a long history \cite{adamopoulou2020chatbots,shawar2007chatbots}, are programs that use Natural Language Processing (NLP) to understand human language through text or voice \cite{khanna2015study,may2022security,motger2022software}. Recent studies show that mental health apps incorporating AI chatbots are rapidly gaining popularity \cite{biswas2025data}. For instance, apps such as ``Wysa: Anxiety, therapy chatbot'' allow users to enter queries via text and receive responses tailored to their input. These chatbots also hold the potential to act as virtual doctors or nurses, offering affordable, round-the-clock care and support \cite{li2023security}.

However, security and privacy concerns have already been raised around health chatbots, yet primarily in the context of scoping and systematic literature reviews \cite{surani2022understanding,may2022security}. More recent research has examined only limited aspects, such as data handling practices (based on Google Play's Data Safety section) and the privacy policies of mental health chatbot apps \cite{biswas2025data}. The most closely related work to ours (see Section \ref{RW}) only empirically analyzes AI-based chatbots, i.e., general-purpose AI chat assistants, without specifically addressing medical and health chatbots. Hence, based on the current gap of evidence on the security and privacy of mHealth chatbots, we developed the following research questions (RQs) to guide our research project:

\begin{itemize}
    \item[] \textbf{RQ1:} What are the current security and privacy risks of selected AI-based mHealth chatbots?
    \item[] \textbf{RQ2:} Based on these findings, what actionable recommendations can be made for developers of the mHealth chatbots?
\end{itemize}

To address the RQs, we followed a three-phase security and privacy analysis process of \textit{manual, static code, and dynamic analyses}. Specifically, we utilize the Mobile Security Framework (MobSF)\footnote{MobSF: \url{https://github.com/MobSF/Mobile-Security-Framework-MobSF}} tool to perform the security analysis from both static and dynamic perspectives. In addition, we manually assess the \texttt{AndroidManifest.xml} files to map requested permissions and evaluate the privacy policies for compliance with Google Play requirements \cite{google}. The latter examines several key aspects, such as secure data handling practices, revealing the best practices to protect users' sensitive data. 

As a result, this study presents the first comprehensive security and privacy analysis of AI-based mHealth chatbots, bridging the current evidence gap and offering a practical set of actionable recommendations to guide developers in enhancing the protection of sensitive health data. These contributions will benefit key stakeholders developing these apps, including mHealth developers, security testers, and privacy engineers. Nevertheless, these insights extend beyond chatbot-based apps, offering value to the broader mHealth ecosystem, which faces similar vulnerabilities.

%% file: Sections/related_work.tex
\section{Related Work} \label{RW}

Despite increasing concerns \cite{ciesla2024ai}, recent research on the security and privacy of health chatbots shows that the area remains underinvestigated \cite{may2022security,li2023security}. Nevertheless, some studies still offer insights closely related to our topic.

Bao et al. \cite{bao2024evaluating} focused on network traffic analysis, assessing third-party usage, and data-sharing behavior of AI-based medical chatbots through dynamic analysis of 91 apps for Android. They also employed ChatGPT for in-depth analysis of how data is shared and which apps transmit data to external servers. While informative, we argue that this study does not constitute a comprehensive security and privacy analysis, as it does not identify or document specific vulnerabilities or privacy risks. Instead, it highlights behaviours that may lead to risks without demonstrating concrete violations. Nonetheless, their third-party analysis raises concerns about potential data sharing with external services.

The work of Biswas et al. \cite{biswas2025data} investigated the data retention policies, regulatory compliance, and discrepancies between the apps’ Data Safety sections and their privacy policies, focusing on mental health chatbot apps. They evaluated ten apps with over 50,000 downloads, five of which overlap with our study. Their findings showed that the apps request excessive sensitive permissions, with \textit{Yana: Your Emotional Companion} requesting the most, followed by \textit{Wysa: Anxiety, therapy chatbot} and \textit{Youper - CBT Therapy Chatbot}. They further reported that \textit{Ada - Check Your Health}, \textit{Youper}, and \textit{Yana} collect the most user data, while \textit{Ada - Check Your Health} shares the most with third parties. Notably, most apps did not specify clear data retention policies, except for \textit{Ada - Check Your Health}. 

In contrast, our research adopts a broader penetration testing methodology and scope. Technically, we combine static, dynamic, and manual analyses to provide an in-depth assessment of security and privacy. This includes network traffic analysis, permission mapping through \texttt{AndroidManifest.xml}, and evaluation of the privacy policy for compliance with Google Play requirements concerning user data. Moreover, while Bao et al. \cite{bao2024evaluating} examined apps that are largely general-purpose AI chat assistants (e.g., \textit{ChatGPT, ChatGen - AI Chatbot \& Writer, Poe - Fast AI Chat}), with only one mental health app (\textit{Wysa: Anxiety, therapy chatbot}), and Biswas et al. \cite{biswas2025data} limited their scope to mental health chatbots, our study exclusively analyzes a broader category of AI-based mHealth chatbots, including medical, mental, and fitness apps.

%% file: Sections/methodology.tex
\section{Methodology}\label{methods}

As mentioned, our methodology is structured around three key phases: manual inspection, static code analysis, and dynamic analysis. These three analyses are commonly used for assessing mobile app security and privacy. Previous work has employed a combination of two (e.g., static and dynamic \cite{forsberg2024security} or manual and static \cite{hatamian2021privacy}), or all three phases (e.g., \cite{papageorgiou2018security,zhao2020security}) as they provide complementary insights. The following subsections outline the data collection process, security and privacy analyses, and the supporting tools (MobSF, Google Play scraper, and APKTool).


\subsection{Data Collection}

Our study focuses on AI-based mHealth chatbots for Android, targeting apps with a large userbase (100,000+ to 10,000,000+ users). We also included lesser-downloaded apps ($\leq$ 100,000), as even these may expose hundreds of users in the event of a data breach, loss of personal data control, or misuse.  We retrieved apps from the Google Play Store using \texttt{google-play-scraper}, a Node.js module\footnote{Google Play Scraper: \url{https://pypi.org/project/google-play-scraper/}}. The automated search, conducted on February 8, 2025, employed the queries \textit{mobile health chatbot}, \textit{medical chatbot}, and \textit{fitness \& wellness chatbot}, returning a total of 81 apps. Since Google Play limits \texttt{google-play-scraper} results to 30 apps, we conducted four separate searches using each query individually and a combined query, without using any country or region restrictions.

To complement the automated search, we conducted a manual search on March 5, 2025, using the keyword \textit{health chatbots}. This allowed us to identify additional top-downloaded apps not captured earlier, ensuring a more diverse and representative sample. 

Before installation, we manually reviewed the retrieved apps to ensure their eligibility. We included apps that fell into the category of mHealth chatbots, including ``medical'' and ``fitness \& wellness'' chatbots, and excluded those that did not provide chatbot functionality. Apps requiring payment for basic functionality were excluded, although paid apps were included if their basic features were accessible without in-app purchases (e.g., \textit{Wysa: Anxiety, therapy chatbot}).

In total, we selected $n = 16$ apps, summarized in Table \ref{tab:analysed_apps}. Due to legal constraints, the specific names and identifiers of the apps cannot be disclosed; however, the countries of origin of the respective chatbot providers are reported as follows: \textit{App1, App15 - Germany; App2 - Mexico; App3, App4, App13 - US; App5 - Poland; App6 - South Korea; App7, App12, App16 - Unspecified; App8, App11 - India; App9 - Japan; App10 - UK; App14 - Pakistan.}

We obtained and decompiled the APKs using APKTool\footnote{APKTool: \url{https://github.com/iBotPeaches/Apktool}}, a tool used for compiling and decompiling APKs \cite{sanders2021comparison} to retrieve the APK files, such as the \texttt{AndroidManifest.xml} and \texttt{smali}. In addition, we collected and archived the apps' privacy policies for subsequent analysis.

\input{Tables/analysed_apps}

\subsection{Manual Inspection}

In phase 1, we conducted two separate analyses:
(i) a Manifest permission mapping, and (ii) a privacy policies analysis to identify discrepancies with the requested permissions, particularly dangerous permissions, and assess compliance with the Google Play policy. 

\subsubsection{Manifest Permission Mapping}
We mapped permissions declared in the app set, within each app's \texttt{AndroidManifest.xml} file. These permissions define what resources or system actions an app can access, falling into four categories as defined by Android developers\footnote{Android's permission categories: \url{https://developer.android.com/guide/topics/permissions/overview}}, i.e., \texttt{Normal}, \texttt{Dangerous}, \texttt{Signature}, and \texttt{Special}. 

We also emphasize \texttt{dangerous} permissions (i.e., high-risk), as they grant access to sensitive data or features and pose potential privacy and security risks in AI-based mHealth chatbots. Furthermore, this mapping serves as a foundation for a privacy policy analysis, enabling us to identify discrepancies between the dangerous permissions requested and the disclosures provided to users.

\subsubsection{Assessing Compliance with Google Play Policy}\label{pp}

Privacy policies are crucial for users to understand how their data is handled, often influencing decisions \cite{jensen2004privacy} about whether to use a particular app. A privacy policy is defined as a \textit{``statement or legal document that gives information about the ways an app provider collects, uses, discloses, and manages users' data''} \cite{hatamian2020engineering}. For health-related apps, such transparency is especially critical given the sensitivity of the data involved. 

Google Play requires apps that process personal and sensitive user data to provide a privacy policy, beyond the \textit{Data Safety} section, and compliant with data protection laws \cite{google}. Specifically, privacy policies must disclose:
\begin{itemize}
    \item \textit{developer information and a privacy point of contact or a mechanism to submit inquiries;}
    \item \textit{disclosing the types of personal and sensitive user data your app accesses, collects, uses, and shares; and any parties with which any personal or sensitive user data is shared;}
    \item \textit{secure data handling procedure for personal and sensitive user data;}
    \item \textit{the developer's data retention and deletion policy;} and, 
    \item \textit{clear labeling as a privacy policy (for example, listed as ``privacy policy'' in the title).}
\end{itemize}

The privacy policy must also be accessible via a publicly accessible URL without any geographic restrictions, in HTML format, be non-editable by users, and stored in the app \cite{google}. As such, this study evaluates the identified mHealth chatbots against these criteria to determine how much they comply with Google Play's requirements.

\subsection{Static code \& Dynamic Analyses}

In phase 2, we conducted the static code analysis of the apps using MobSF. This tool, in addition to being popular \cite{kouliaridis2023assessing,iwaya2023privacy}, is widely used for assessing Android, iOS, and Windows mobile apps \cite{khan2024android} and supports analysis of APKs. MobSF supports tests such as app permissions, trackers, hardcoded secrets, network, certificate, code, and manifest analyses \cite{forsberg2024security}. The reports generated provide a snapshot of each app's security posture based on these tests.

Unlike static analysis, which examines the code without execution, dynamic analysis (phase 3) evaluates the app during runtime. This enables the identification of vulnerabilities that may not be detectable statically \cite{bao2024evaluating}. As such, we used MobSF to conduct dynamic assessments, which include runtime monitoring, network traffic analysis, and interactive instrumented testing such as logging data. 

Finally, using the results from both static and dynamic analyses, we assessed whether discrepancies exist between the claims made in privacy policies and the apps' actual behaviour. This includes evaluating secure data handling and data-sharing practices. Such discrepancy analysis ensures that privacy policies are not merely formalities written to comply with regulations, but rather reflect an app's actual data practices.

\subsection{Ethical approval} 
This research is part of the project ``Comprehensive Quality Assessment of Mobile Apps,'' whose ethical application was approved by the local ethics committee at Karlstad University (diary number: HNT 2023/795). The security and privacy analysis was conducted solely to understand how AI-based mHealth chatbots handle sensitive data and implement security practices. We do not collect any personal data, tamper with servers, or exploit any vulnerabilities. Findings were responsibly disclosed to the developers using their privacy policies contacts.

%% file: Tables/analysed_apps.tex
\begin{table}[h]
\scriptsize
    \centering
    \caption{List of AI-based mHealth chatbots selected and classified based on the number of installs.}
    \label{tab:analysed_apps}
        \centering
        \begin{tabular}{|l|c||l|c|}
            \hline
            \textbf{Apps} & \textbf{Installs} & \textbf{Apps} & \textbf{Installs} \\
            \hline
            App1, App2 & 10M+ & App11, App12, App13 & 10K+ \\
            App3, App4 & 1M+ & App14, App15 & 5K+ \\
            App5 & 500K+ & App16 & 1K+ \\
            App6, App7, App8, App9, App10 & 100K+ & & \\
            \hline
        \end{tabular}
\end{table}

%% file: Sections/results.tex
\section{Findings}\label{results}

\subsection{Manual Analysis}
 
\subsubsection{Manifest Permission Mapping}

In total, the 16 apps request 62 unique permissions; of these, 47 (75.8\%) categorized as \texttt{normal}, 11 (17.7\%) as \texttt{dangerous}, 3 (4.8\%) as \texttt{signature}, and 1 (1.6\%) as \texttt{special} permission levels. However, focusing only on the dangerous permissions, Table~\ref{tab:dangerous} shows that the most frequent permission is \texttt{POST\_NOTIFICATIONS}, followed by \texttt{WRITE\_EXTERNAL\_STORAGE}, \texttt{READ\_EXTERNAL\_STORAGE}, \texttt{CAMERA}, and \texttt{RECORD\_AUDIO}.

\input{Tables/dangerous}


In fairness, most permissions requested align with the apps' intended functionalities, such as \texttt{POST\_NOTIFICATIONS} required to deliver notifications or updates to the app. Similarly, the \texttt{CAMERA} permission has explicit purposes, e.g., \textit{App8} uses it for biometric information (i.e., face scanning), \textit{App6} for biometrics and capturing photos of medical reports, and \textit{App4} for services provided to users on demand. However, this justification is absent in some cases, i.e., \textit{App3}, \textit{App7}, \textit{App14}, and \textit{App15}, which request camera access but do not explicitly explain their purposes in their privacy policies. Some apps may plausibly use the camera (e.g., video consultations in \textit{App15}), but their policies' lack of clear justification raises transparency concerns, and the access to high-risk data can be considered privacy invasive by users.

We also observed the request of \texttt{WRITE\_EXTERNAL\_STORAGE} and \texttt{READ\_EXTERNAL\_STORAGE}, both of which are dangerous as they allow apps to read and modify files in external storage. In fact, these permissions were deprecated since APIs level 30 and level 33, respectively \footnote{See: \url{https://developer.android.com/reference/android/Manifest.permission#WRITE_EXTERNAL_STORAGE} and \url{https://developer.android.com/reference/android/Manifest.permission#READ_EXTERNAL_STORAGE}}. Instead, developers should request \texttt{READ\_MEDIA\_IMAGES} and \texttt{READ\_MEDIA\_VIDEO} to access media files. Consequently, using deprecated permissions suggests a lack of awareness of up-to-date Android developer storage policies, which can lead to compatibility issues with newer API levels and increase security and privacy risks.

\subsubsection{Assessing Compliance with Google Play Policy}

Further analysis of the app's privacy policy against the Google Play policy also generated worrying results for some apps (see Table \ref{tab:policy}).

\input{Tables/policy_analysis}

Our findings indicate that \textit{App7} is the least compliant with the Google Play policy. Unlike most other apps, it not only failed to provide a privacy policy in the Play Console but was also missing both retention and deletion policies, a dedicated section on secure data handling, and any clear developer or inquiry mechanism. Such shortcomings violate fundamental privacy principles, including transparency, accountability, and user control over personal data. Similarly, \textit{App12} also has significant deficiencies, some of which align with those of \textit{App7}.

\subsection{Static Code \& Dynamic Analysis}

\subsubsection{Static Code Analysis}

Table~\ref{tab:issues} summarizes the main issue categories identified for each app, of which the most prevalent were third-party trackers (15/16), Manifest issues (12/16), and code issues (9/16). 
As follows, we further detail our findings for these issue categories.

\input{Tables/issues}

\subsubsection{Third-Party Trackers Detected}

The tracker analysis varies across apps, with \textit{App16} alone accounting for 15 third-party trackers (22.7\% of all trackers identified), while others have significantly fewer (see Tab~\ref{tab:static_tracker}). The Google Firebase Analytics tracker was the most frequently used, widely employed for analytics and performance monitoring as outlined in Table \ref{tab:tracker}. Notable differences emerged between declared privacy policies and actual tracker usage. For example, \textit{App16}'s policy only acknowledges the use of Stripe (for payments) and Google Play services (for app functionality, user authentication, and analytics). Yet, the static analysis detected a broader range of trackers, suggesting incomplete disclosure, which raises compliance concerns and erodes users' trust. Conversely, \textit{App5} declared the use of Google Analytics in its privacy policy, but this did not appear in the analysis.

\input{Tables/static_trackers}

\input{Tables/trackers}


\subsubsection{Issues Detected in the Manifest}

Table \ref{tab:manifest_analysis} shows the manifest-related issues, with two problems standing out. First, 62.5\% of the apps were installable on outdated and unsupported Android versions (e.g., Android 7.0 or lower), exposing users to known vulnerabilities that are exploitable if the device lacks security patches. For example, on devices running Android versions before 6.0, runtime permissions were granted at installation time and could not be revoked by users, thereby reducing control over access to high-risk data and resources\footnote{\url{https://developer.android.com/about/versions/marshmallow/android-6.0-changes}}. Second, 25\% of apps had cleartext traffic enabled, allowing unencrypted HTTP communication, creating risks of health data interception if network traffic is compromised. 
Another high-severity issue was identified in \textit{App3} (see Listing \ref{lst:intent}), where the app link configuration included a custom URL scheme instead of being limited to \texttt{http} or \texttt{https}\footnote{\mbox{\url{https://developer.android.com/training/app-links/verify-android-applinks}}}. The app links are web links \textit{``that use the HTTP and HTTPS schemes and contain the autoVerify attribute''}\footnote{\url{https://developer.android.com/training/app-links}}. This misconfiguration can break the auto-verification process and expose the app to security risks, including hijacking. These findings highlight that misconfigurations can significantly reduce the security posture of mHealth apps even at the manifest level.

\input{Tables/manifest_analysis}

\begin{figure}[h!]
    \centering
    \label{youper}
    \begin{minipage}{\linewidth}
        \begin{lstlisting}[language=json, caption={JSON representation of AndroidManifest.xml snippet for App3 showing the app's intent filters.}, label={lst:intent}]
<intent-filter android:autoVerify="true">
<action android:name="android.intent.action.VIEW"/>
<category android:name="android.intent.category.DEFAULT"/>
<category android:name="android.intent.category.BROWSABLE"/>
<data android:scheme="@string/custom_url_scheme"/>
<data android:host="****.**" android:scheme="https"/>
</intent-filter>
        \end{lstlisting}
    \end{minipage}
\end{figure}

\subsubsection{Issues Detected in the Code}

The code analysis (Table~\ref{tab:code_analysis}) revealed several vulnerabilities mapped to the Common Weakness Enumeration (CWE), the Open Worldwide Application Security Project (OWASP) top 10, and the OWASP Mobile Application Security Verification Standard (MASVS), some of which could have serious security implications. While specific findings may initially appear minor, such as warnings about insecure random number generators (\textit{App1}), these can still expose apps to significant risks if left unaddressed. However, focusing on high-severity issues, the most prevalent concern was the enabling of Remote WebView debugging, found in 43.7\%, which could allow attackers to inspect and manipulate content via Android Debug Bridge (adb)\footnote{\url{https://developer.android.com/reference/android/webkit/WebView.html\#setWebContentsDebuggingEnabled(boolean)}}. Cryptographic weaknesses were also found, where 31.2\% of the apps used CBC mode with PKCS5/PKCS7 padding, which is vulnerable to padding oracle attacks. Also, 12.5\% relied on ECB mode, which is insecure due to ciphertext manipulation attacks \cite{vaudenay2002security}.

\input{Tables/code_analysis}

\subsubsection{Network Security Issues}

Three apps were flagged with high-severity issues as outlined in Table~\ref{tab:issues}. In all cases, the \texttt{network\_security\_config} was insecurely configured, allowing cleartext traffic, unencrypted HTTP connections, and undermining Android's default protections (introduced from API level 28\footnote{\url{https://developer.android.com/privacy-and-security/security-config}}). For \textit{App1} and \textit{App9} (see example in Listing \ref{lst:network-config-1}), the insecure configurations primarily reference emulator network addresses (i.e., \texttt{10.0.2.2}, \texttt{10.0.3.2}, and \texttt{127.0.0.1}), typically used for testing during development\footnote{\url{https://developer.android.com/studio/run/emulator-networking}}. While not inherently dangerous, enabling cleartext traffic in production is a poor practice, as it risks exposing sensitive health data. 

\textit{App8} presented a more concerning case (see Listing \ref{lst:network-config-2}). Besides the localhost entry, its configuration explicitly permitted cleartext traffic for its own \texttt{api-server} and an Amazon-owned domain. The lack of HTTPS enforcement could expose users' health information to data leakage or interception. Although Amazon EC2\footnote{\url{https://docs.aws.amazon.com/AWSEC2/latest/UserGuide/concepts.html}} may be necessary, developers should configure secure connections and explicitly disable cleartext traffic to comply with Android security requirements.

\begin{figure}[h]
    \centering
    \begin{minipage}{\linewidth}
        \begin{lstlisting}[language=json, caption={JSON representation of network\_security\_config for App1}, label={lst:network-config-1}]
<?xml version="1.0" encoding="utf-8"?>
<network-security-config>
    <domain-config cleartextTrafficPermitted="true">
        <domain includeSubdomains="true">10.0.2.2</domain>
        <domain includeSubdomains="true">10.0.3.2</domain>
        <domain includeSubdomains="true">localhost</domain>
    </domain-config>
</network-security-config>
        \end{lstlisting}
    \end{minipage}
\end{figure}

\begin{figure}[h]
    \centering
    \begin{minipage}{\linewidth}
        \begin{lstlisting}[language=json, caption={JSON representation of network\_security\_config for App8}, label={lst:network-config-2}]
<?xml version="1.0" encoding="utf-8"?>
<network-security-config>
    <domain-config cleartextTrafficPermitted="true">
        <domain includeSubdomains="true">api-server.****</domain>
        <domain includeSubdomains="true">localhost</domain>
        <domain includeSubdomains="true">****</domain>
    </domain-config>
</network-security-config>
        \end{lstlisting}
    \end{minipage}
\end{figure}

\subsubsection{Firebase Database Misconfigurations}

Three apps (\textit{App3}, \textit{App8}, and \textit{App14}) suffered from Firebase database misconfigurations, where the databases were openly accessible without authentication (i.e., only required access to the URL link). This exposes the associated \texttt{JSON} files, which may contain sensitive app or user data. For example, in the case of \textit{App14}, the exposed database revealed details including API keys, chatbot prompts and responses, and device-related information. While we cannot determine whether the exposed data originates from real users or test accounts, such exposure could constitute a significant data breach that is easily exploitable by malicious actors.

\subsubsection{Dynamic analysis}

Out of the 16 apps, only 13 could be subjected to dynamic analysis. Three apps (\textit{App5}, \textit{App4}, and \textit{App9}) could not be emulated due to anti-VM or root detection mechanisms. Among the 13 analyzed, seven (\textit{App12}, \textit{App15}, \textit{App13}, \textit{App14}, \textit{App11}, \textit{App6}, and \textit{App2}) experienced frequent crashes during testing, likely due to root detection measures or instrumentation environment conflicts. Nonetheless, we could still conduct the analyses and extract meaningful runtime data. As follows, we present the key findings, focusing on network traffic patterns and behavioral analysis, also emphasizing third-party trackers detected during execution.

\subsubsection{HTTP(S) Traffic Analysis}

Although most apps transmitted data securely over HTTPS, three apps (\textit{App7}, \textit{App12}, and \textit{App14}) initially failed the cleartext test. However, upon further manual investigation, we realized it was not anything concerning as these apps were transmitting data over HTTPS. 



A more critical finding was that these apps transmitted personally identifiable information (PII), such as \texttt{userIDs} and \texttt{deviceIDs}, even when the data was encrypted over HTTPS. For example, \textit{App1} shared sensitive information, including names, email addresses, and location, with Braze, a service used for push notifications. Although this practice is disclosed in the app's privacy policy, the extent of the data transmitted raises concerns regarding compliance with the principle of data minimization. If exposed, whether maliciously or accidentally, such data can directly identify individuals.

\subsection{Trackers Detected During Runtime Execution}

Dynamic analysis confirmed many of the identified trackers during static analysis (i.e., 9 apps were consistent), and further revealed trackers that were only shown at runtime (i.e., 4 apps were not consistent). For example, \textit{App7} had six trackers detected in the static analysis but 14 at runtime, highlighting the importance of capturing runtime behaviours.
The newly detected trackers included Adform, Amobee, Criteo, Integral Ad Science, Open X, PubNative, Quantcast, Smart, and Taboola. From a privacy perspective, these runtime-observed trackers indicate additional exposure of sensitive data that may not be fully disclosed in the apps' privacy policies. For instance, some trackers pose higher risks to users as they use data for profiling/advertisement, e.g., OpenX's purpose is to provide analytics and programmatic advertising by tracking user behavior. Similarly, Quantcast offers profiling and analytics. It is based on AI to provide advertisements to its users\footnote{\url{https://www.quantcast.com/advertiser}}.

\subsection{Acknowledgements of Vulnerability Disclosures}

We responsibly disclosed the identified vulnerabilities to the app developers. Four developers acknowledged our findings and indicated that they would assess them internally. For example, the developer of \textit{App13} explained that the WebView debugging issue originated from a deprecated third-party SDK that would be removed in an upcoming update. However, we have not followed up to verify whether these vulnerabilities have been addressed since then.

%% file: Tables/dangerous.tex
\begin{table}[h]
\scriptsize
  \centering
  \caption{Overview of frequent dangerous permissions requested}
    \begin{tabular}{|c|p{0.45\textwidth}|p{0.45\textwidth}}
    \hline
    \textbf{Permission count} & \textbf{Permission} \\
    \hline
    13 & POST\_NOTIFICATIONS \\
     8 & WRITE\_EXTERNAL\_STORAGE \\
     7 & READ\_EXTERNAL\_STORAGE \\
     6 & CAMERA \\
     5 & RECORD\_AUDIO \\
     2 & READ\_MEDIA\_IMAGES \\
     2 & ACCESS\_COARSE\_LOCATION \\
    \hline
    \end{tabular}%
  \label{tab:dangerous}%
\end{table}%

%% file: Tables/policy_analysis.tex
\begin{table}[h!]
\scriptsize
  \centering
  \caption{Policy analysis based on each app's privacy policy against Google Play's policy on Users' data. A checkmark indicates compliance with a given criterion as outlined in the privacy policy, per Google Play's requirements, while a cross indicates missing information. The criteria are detailed in Subsection~\ref{pp}.}
  \renewcommand{\arraystretch}{1.2} 
  \begin{tabular}{|l|c|c|c|c|c|c|c|c|c|c|c|c|c|c|c|c|}
    \hline
    \textbf{Criterion} & 
    \rotatebox{90}{App1} & 
    \rotatebox{90}{App2} & 
    \rotatebox{90}{App3} & 
    \rotatebox{90}{App4} & 
    \rotatebox{90}{App5} & 
    \rotatebox{90}{App6} & 
    \rotatebox{90}{App7} & 
    \rotatebox{90}{App8} & 
    \rotatebox{90}{App9} & 
    \rotatebox{90}{App10} & 
    \rotatebox{90}{App11} & 
    \rotatebox{90}{App12} & 
    \rotatebox{90}{App13} & 
    \rotatebox{90}{App14} & 
    \rotatebox{90}{App15} & 
    \rotatebox{90}{App16} \\
    \hline
    Developer Info             & \checkmark & \checkmark & \checkmark & \checkmark & \checkmark & \checkmark & \xmark & \checkmark & \checkmark & \checkmark & \checkmark & \xmark & \checkmark & \checkmark & \checkmark & \checkmark \\
    Privacy Contact            & \checkmark & \checkmark & \checkmark & \checkmark & \checkmark & \checkmark & \checkmark & \checkmark & \checkmark & \checkmark & \checkmark & \checkmark & \checkmark & \checkmark & \checkmark & \checkmark \\
    Data Collected             & \checkmark & \checkmark & \checkmark & \checkmark & \checkmark & \checkmark & \checkmark & \checkmark & \checkmark & \checkmark & \checkmark & \checkmark & \checkmark & \checkmark & \checkmark & \checkmark \\
    Data Sharing \& Third Parties & \checkmark & \checkmark & \checkmark & \checkmark & \checkmark & \checkmark & \checkmark & \checkmark & \checkmark & \checkmark & \checkmark & \checkmark & \checkmark & \checkmark & \checkmark & \checkmark \\
    Secure Data Handling       & \checkmark & \checkmark & \checkmark & \checkmark & \checkmark & \xmark & \xmark & \checkmark & \checkmark & \checkmark & \checkmark & \checkmark & \checkmark & \checkmark & \checkmark & \checkmark \\
    Retention Policy           & \checkmark & \checkmark & \checkmark & \checkmark & \checkmark & \checkmark & \xmark & \checkmark & \xmark & \checkmark & \checkmark & \xmark & \checkmark & \checkmark & \checkmark & \xmark \\
    Deletion Policy            & \checkmark & \checkmark & \checkmark & \checkmark & \checkmark & \checkmark & \xmark & \checkmark & \xmark & \checkmark & \checkmark & \xmark & \checkmark & \checkmark & \checkmark & \checkmark \\
    Privacy Policy Label       & \checkmark & \checkmark & \checkmark & \checkmark & \checkmark & \checkmark & \checkmark & \checkmark & \checkmark & \checkmark & \checkmark & \checkmark & \checkmark & \checkmark & \checkmark & \checkmark \\
    Privacy Policy Accessibility & \checkmark & \checkmark & \checkmark & \checkmark & \checkmark & \checkmark & \xmark & \checkmark & \checkmark & \checkmark & \checkmark & \xmark & \checkmark & \checkmark & \checkmark & \checkmark \\
    \hline
  \end{tabular}
  \label{tab:policy}
\end{table}

%% file: Tables/issues.tex
\begin{table}[h!]
\scriptsize
  \centering
  \caption{Categories of prevalent security issues and affected apps identified through static code analysis}
    \begin{tabular}{|p{3.5cm}|p{8.5cm}|}
    \hline
    \textbf{Issue category} & \textbf{Affected Apps}  \\
    \hline
    Third-Party Trackers Detection & \textit{(All Apps except App5)}  \\
    Manifest issues & \textit{App2, App3, App4, App6, App7, App8, App10, App11, App12, App14, App15, App16} \\
    Code issues & \textit{App2, App6, App7, App8, App9, App13, App14, App15, App16}  \\
    Network Security & \textit{App1, App8, App9}  \\
    Firebase Database Misconfigurations & \textit{App3, App8, App14}  \\
    \hline
    \end{tabular}%
  \label{tab:issues}%
\end{table}%

%% file: Tables/static_trackers.tex
\begin{table}[h]
\scriptsize
  \centering
  \caption{Number of Third-Party Trackers per App Detected During Static Analysis}
    \begin{tabular}{|c|p{0.1\textwidth}||c|p{0.1\textwidth}|}
    \hline
    \textbf{Number of trackers} & \textbf{Apps} & \textbf{Number of trackers} & \textbf{Apps}\\
    \hline
     3 & App1 & 4 & App9 \\
     6 & App2 & 4 & App10 \\
     3 & App3 & 1 & App11 \\
     3 & App4 & 2 & App12 \\
     0 & App5 & 5 & App13 \\
     3 & App6 & 3 & App14 \\
     6 & App7 & 2 & App15 \\
     6 & App8 & 15 & App16 \\
    \hline
    \end{tabular}%
  \label{tab:static_tracker}%
\end{table}%

%% file: Tables/trackers.tex
\begin{table}[h!]
\scriptsize
  \centering
  \caption{Trackers Identified Within the App Set (N: Number of Apps)}
    \begin{tabular}{|c|p{0.45\textwidth}||c|p{0.45\textwidth}|}
    \hline
    \textbf{N} & \textbf{Trackers} & \textbf{N} & \textbf{Trackers} \\
    \hline
    14 & Google Firebase Analytics & 1 & CleverTap \\
     8 & Google CrashLytics & 1 & OpenTelemetry  \\
     6 & Google AdMob & 1 & AdColony \\
     4 & AppsFlyer & 1 & Facebook Ads \\
     4 & OneSignal & 1 & Inmobi \\
     3 & MixPanel & 1 & Mintegral \\
     3 & Facebook Login & 1 & Pangle \\
     3 & Facebook Share & 1 & Startapp \\
     2 & Facebook Analytics & 1 & Unity3d Ads\\
     2 & IAB Open Measurement & 1 & ironSource \\
     2 & AppLovin - Max and SparkLabs & 1 & Flurry  \\
     1 & Adjust & 1 & Amplitude \\
     1 & Sentry & 1 & Branch \\
    \hline
    \end{tabular}%
  \label{tab:tracker}%
\end{table}%

%% file: Tables/manifest_analysis.tex
\begin{table*}
\scriptsize
  \centering
  \caption{Issues detected from manifest analysis with a high severity}
    \begin{tabular}{|p{1.9cm}|p{10.0cm}|}
    \hline
    \textbf{App} & \textbf{Issue and Description} \\
    \hline

    App3 & App Link assetlinks.json file not found -- \emph{Description:} App Link asset verification URL not found or misconfigured. \\
     \hline

    App4 &  App can be installed on a vulnerable, unpatched Android version Android 4.1-4.1.2, [minSdk=16] -- \emph{Description:} The app can be installed on an older Android version with multiple unfixed vulnerabilities. \\

    \hline

    App6, App14, App15 & App can be installed on a vulnerable, unpatched Android version Android 6.0-6.0.1, [minSdk=23] -- \emph{Description:} The app can be installed on an older Android version with multiple unfixed vulnerabilities. \\
     \hline
    
    App6, App7, App8, App16 & Clear text traffic is enabled for App -- \emph{Description:} The app intends to use cleartext network traffic, such as cleartext HTTP, FTP stacks, DownloadManager, and MediaPlayer. \\
     \hline

    App2, App7, App10, App12 & App can be installed on a vulnerable, unpatched Android version Android 7.0, [minSdk=24] -- \emph{Description:} The app can be installed on an older Android version with multiple unfixed vulnerabilities. \\
    \hline

    App11, App16 & App can be installed on a vulnerable, unpatched Android version Android 5.0-5.0.2, [minSdk=21] -- \emph{Description:} The app can be installed on an older Android version with multiple unfixed vulnerabilities. \\

    \hline
    \end{tabular}%
  \label{tab:manifest_analysis}%
\end{table*}%

%% file: Tables/code_analysis.tex
\begin{table*}[h!]
\scriptsize
  \centering
  \caption{Issues detected from code analysis with a high severity}
    \begin{tabular}{|p{2.10cm}|p{9.70cm}|}
    \hline
    \textbf{App} & \textbf{Issue and Standards} \\
    \hline
    
    App2, App6, App7, App9, App13, App15, App16 & 
    \textbf{Issue:} Remote WebView debugging is enabled \newline
    \textbf{Standards:} \textbf{CWE: CWE-919} Weaknesses in Mobile Applications;
         \textbf{OWASP Top 10:} M1: Improper Platform Usage; \textbf{OWASP MASVS:} MSTG-RESILIENCE-2 \\ \hline
    
    App6, App8, App9, App14, App16 & \textbf{Issue:} The app uses the encryption mode CBC with PKCS5/PKCS7 padding. \newline
    \textbf{Standards:} \textbf{CWE: CWE-649:} Reliance on Obfuscation or Encryption of Security-Relevant Inputs without Integrity Checking; \textbf{OWASP Top 10:} M5: Insufficient Cryptography; \textbf{OWASP MASVS:} MSTG-CRYPTO-3 \\ \hline

    App14, App16 & \textbf{Issue:} The app uses ECB mode in its cryptographic encryption algorithm. \newline
    \textbf{Standards:} \textbf{CWE: CWE-327:} Use of a Broken or Risky Cryptographic Algorithm; \textbf{OWASP Top 10:} M5: Insufficient Cryptography; \textbf{OWASP MASVS:} MSTG-CRYPTO-2 \\ \hline
    \end{tabular}%
  \label{tab:code_analysis}%
\end{table*}%

%% file: Sections/discussion.tex
\section{Discussion} \label{discussion}
\subsection{Permissions and Privacy Policy Transparency}

Dangerous permission analysis against the apps' privacy policies revealed a lack of transparency in certain apps regarding the request for \texttt{CAMERA}. Such an omission raises privacy concerns, as users may grant sensitive permissions without fully understanding the implications, potentially exposing themselves to unnecessary data collection (e.g., surveillance \cite{solove2005taxonomy}). This discrepancy between requested dangerous permissions and stated privacy policies is also observed in other AI-based apps, such as romantic AI chatbot apps \cite{ragab2024trust}, hence suggesting it may be a common issue. Given this, developers of these apps need to be explicit about data collection and provide reasonable justifications as to why the app requires certain features.

Regarding privacy policy compliance, the majority of apps (11/16) adhered to Google Play's policy, demonstrating a clear handling of user data. However, five apps (\textit{App6}, \textit{App7}, \textit{App9}, \textit{App12}, and \textit{App16}) failed to fully comply, particularly by omitting data retention and deletion policies. Although this represents a minority within our sample (31.23\%), a large-scale study of Android apps by Verderame et al. found that only 0.9\% (46/5057) of analyzed apps' privacy policies fully complied with Google Play Guidelines \cite{verderame2020reliability}, suggesting that noncompliance is a widespread issue across the Android app ecosystem. This noncompliance raises concerns as users may lack clarity on how long their data is stored or whether it could be used to train AI models, highlighting the need for developers to provide, for example, explicit retention and deletion procedures. Additionally, the lack of accessible privacy policies, as seen in \textit{Apps 7 and 12}, prevents users from understanding how their data is processed, including details on whom to contact regarding processing activities, which is essential for users to make informed decisions about whether to use a particular app.

\subsection{Miscofigurations, Insecure Coding, and Third-Party Trackers}

The static code analysis revealed several vulnerabilities, as listed in Table \ref{tab:issues}. Manifest analysis revealed that 62.5\% of the apps can be installed on unpatched Android versions vulnerable to known security issues, and 25\% permit cleartext traffic, allowing insecure HTTP transmissions. Additionally, 43.7\% of apps had Remote WebView debugging enabled, shown in Table \ref{tab:code_analysis}. This aligns with findings by \cite{papageorgiou2018security}, who reported that 50\% of the mHealth apps they analyzed had similar vulnerabilities. Furthermore, we observed that 31.2\% used the CBC encryption mode with PKCS5/PKCS7, indicating a lack of security awareness or poor secure coding practices.

These weaknesses significantly increase the attack surface and could lead to the compromise of sensitive health data. For example, one of the mHealth security dimensions, such as confidentiality, \cite{plachkinova2015taxonomy}, when breached through, for instance, a padding oracle attack, could result in unauthorized access to sensitive information and erode user trust. 

The static analysis also highlighted privacy concerns related to third-party trackers. While trackers may have a purpose, they also raise privacy concerns that users may be unaware of profiling and advertising practices due to the general lack of transparency from app providers \cite{razaghpanah2018apps}. Furthermore, we found discrepancies between privacy policy declarations and actual tracker usage, as seen in \textit{App16}, which claimed to use only 2 third-party services, but 15 trackers were identified at runtime. Exposure to such a high number of trackers is particularly worrying for sensitive mHealth chatbots.

\subsection{Insecure Communication and Excessive Data Sharing}

Regarding the traffic analysis, we observed from the logs that the apps transmitted data over HTTPS, indicating that developers are aware of the need to secure sensitive data or attentive to the Google Play requirement for secure handling of personal and sensitive user data \cite{google}. This is a positive finding, as HTTPS ensures that data is encrypted in transit, reducing the risk of interception.
 
We also observed that the apps connected to many third-party services during the analysis. According to Xinyu et al \cite{xinyu2023andetect}, \textit{``third-party libraries have become a prevalent feature, particularly among developers of free applications''}. This is observed with the number of third-party trackers connected to the apps, as observed in both static and dynamic analysis. Most of these trackers are used for profiling users to provide targeted advertisements, while others focus on analytics (e.g., Adform and OpenX). More trackers were also identified during static analysis than during dynamic analysis, for example, \textit{App7} revealed connections to more third parties during runtime, indicating a lack of transparency in data sharing.

\subsection{Recommendations}

In light of these findings, we provide a set of recommendations that developers should follow to align with best practices for privacy and security. To develop these recommendations, we combined our empirical findings with inspiration from domain experts on ensuring the security of mHealth apps, as discussed in \cite{aljedaani2020empirical}.

\begin{itemize}
    \item \textbf{Audit third-party dependencies:} Developers should monitor third-party SDKs to ensure they are not deprecated, thereby avoiding potential vulnerabilities (e.g., App13). Additionally, they should assess them to ensure they do not have hidden trackers that may harm user privacy. Furthermore, developers should rely on well-established, secure libraries, instead of custom or unverified code, as best practice for ensuring app security \cite{aljedaani2020empirical}.

    \item \textbf{Permissions requests:} Where permissions are required, their purposes should be explicitly justified within the privacy policy. Additionally, developers should be mindful of using updated permissions in accordance with the Android developer guidelines. 

    \item \textbf{Attentiveness to Google Play requirements:} Developers need to be fully aware of the guidelines on user data, particularly ensuring that they are transparent about how they handle the data, i.e., comprehensively disclose how they access, collect, use, store, and share user personal and sensitive data. This not only ensures compliance with respective data protection laws, but also trust toward users of the apps. 

    \item \textbf{Adopt proactive security and privacy practices:} We recommend incorporating security and privacy threat modeling (e.g., using STRIDE \cite{shostack2014threat} and LINDDUN \cite{wuyts2020linddun}) during development, complemented by post-deployment analysis to identify and mitigate emerging security and privacy issues \cite{aljedaani2020empirical}. 

    \item \textbf{Apply privacy design strategies \cite{hoepman2014privacy}:} Strategies such as minimize and inform should be applied to data minimization, even when secure protocols are used, and to inform users that ensuring transparency. 
\end{itemize}

Users, on the other hand, need to assess whether an app has a privacy policy that outlines the data collection practices and secure handling measures, including information about the developer. However, while this is the case, privacy policies need to be simplified for users to grasp \cite{haggag2022large}. At the same time, if an app requests permissions to features or data that have not been justified, the user should spot the app as not following good privacy practice.

%% file: Sections/limitations.tex
\section{Limitations} \label{limits}

This study has limitations that should be acknowledged. First, after the automatic search, the app selection process followed a manual search in Google Play using the term \textit{health chatbots}. While this approach captured widely downloaded apps, it may have excluded relevant apps discoverable under different keywords such as \textit{mHealth chatbot} or \textit{mobile health chatbot}. Although this might have returned overlapping results, the possibility of identifying new ones would have emerged. 

Both static and dynamic analyses were performed using the MobSF tool. While the tool is popular for performing security analysis, it is prone to false positives \cite{papageorgiou2018security}. Hence, during the study, we relied on expert knowledge and sought to manually verify specific reported issues. This helped confirm their plausibility. Nevertheless, we acknowledge that using other tools to confirm the issues could have been a viable alternative.

%% file: Sections/conclusion.tex
\section{Conclusion} \label{conclude}
Leveraging AI for mHealth apps helps to bridge healthcare gaps, benefiting underserved communities, and providing more readily available access. When properly used, AI chatbots can help to support people's health, leading to improved outcomes. However, due to the sensitive context of such apps and the predominance of an advertising-based business model for revenue generation, such solutions can end up causing more harm than good.

Through a comprehensive security and privacy analysis of AI-based mHealth chatbots, this study highlights several security and privacy challenges. They include non-compliance with Google Play policy requirements and security vulnerabilities that malicious actors could exploit. Addressing these issues is essential for enhancing user trust, safeguarding sensitive data, and ensuring compliance with regulatory standards. To this end, we have outlined best practice recommendations to guide developers in strengthening the security and privacy of AI chatbot-based mHealth apps. 

Looking ahead, we plan to extend this work by looking into the algorithmic transparency of these apps. Particularly, we aim to evaluate whether these apps provide an explainability statement that clarifies how the chatbot works and why it is used. Furthermore, we will assess whether they run their own model or utilize external models to provide services. Additionally, we aim to conduct usability testing of these apps, focusing on their conversational interfaces and natural language interactions, gathering insights into how these apps are used. Lastly, we would conduct a follow-up study to see if the issues identified and reported were fixed.

%% file: Sections/credits.tex
\begin{credits}
\subsubsection{Acknowledgments.} This work was supported in part by the Knowledge Foundation of Sweden (KKS), Region Värmland (Grant: RUN/230445), and the European Regional Development Fund (ERDF) (Grant: 20365177) in connection with the DHINO 2 project, and Vinnova (Grant: 2018-03025) via the DigitalWell Arena project.

\subsubsection{\discintname}
The authors declare that they have no known competing financial interests or personal relationships that could have appeared to influence the work reported in this paper.
\end{credits}